\documentstyle[11pt,iau185,twoside,epsf]{article}

\begin{document}
\title{A New Way to Resolve Cepheid Binaries
\altaffilmark{1}}
\author{Nancy Remage Evans}
\affil{Smithsonian Astrophysical Observatory, Cambridge, 
MA, USA }
\author{Derck Massa}
\affil{Emergent IT, MD, USA }
\altaffiltext{1}{Based on observations made with the 
NASA/ESA Hubble
Space Telescope, obtained at the Space Telescope Science 
Institute,
which is operated by the Association of Universities for 
Research in
Astronomy, Inc. under NASA Contract No. NASA-26555}

\begin{abstract} We have measured the centroid shift of low 
resolution 
HST FOC spectra as the dominant star changes from the 
Cepheid 
to the hot companion. With this approach we have resolved 
the AW Per 
system and marginally resolved the U Aql system.
\end{abstract}

\keywords{Stars: oscillations, Stars: variables: slowly 
pulsating B stars, 
Line: profiles, Binaries: spectroscopic}

\section{Introduction}

Spatially resolving Cepheid binary systems is the first 
step in measuring 
the inclination of the orbit. A successful resolution 
means that ultimately both 
the mass and distance can be determined from first 
principles. 
This provides cosmologically important distances, and also 
benchmarks 
for stellar evolution calculations.

\begin{figure}
\begin{center}

\mbox{\epsfxsize=0.9\textwidth\epsfysize=0.9\textwidth\epsfbox[65 
80 550 690]{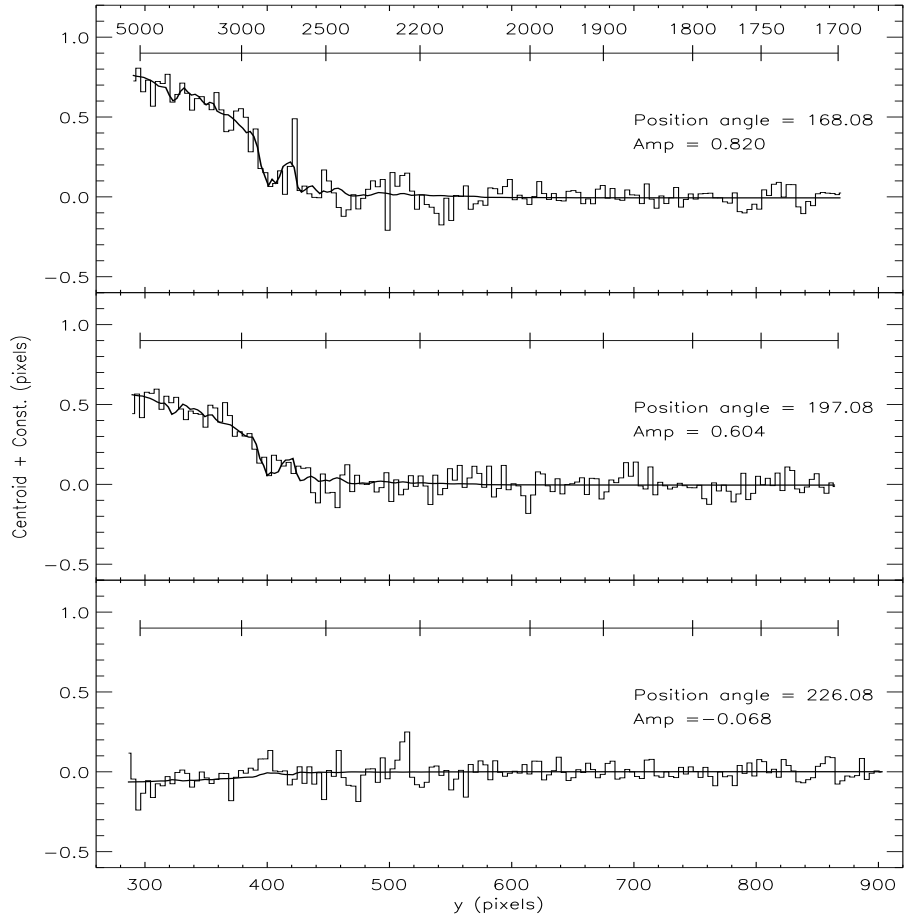}}
\caption{The location of the centroid of the AW Per spectra 
shown for 3 roll
angles (specified on the right) in the 3 panels. Both 
coordinates are in
pixels, but the wavelength scale is given at the top of the 
figure. The data are
shown as histograms; the solid line is the model derived 
from the flux
distribution. }
\end{center}\end{figure}

\section{Observations} 

We have developed a new approach to resolving Cepheid 
binaries using the Hubble
Space Telescope Faint Object Camera (HST FOC) in the 
objective prism mode. The
basic approach was first put forward by Massa \& Endal 
(1987a,b). In
Cepheid binary systems, bright companions will be hot main 
sequence stars. In
this case, the Cepheid dominates in the visible part of the 
spectrum, however it
contributes very little flux to the ultraviolet spectrum, 
where the hot
companion dominates. The two components may be unresolved 
in an image. In the
low resolution FOC spectra, however, the location of the 
centroid of the
spectrum on the detector will shift as a function of 
wavelength as the star
producing the spectrum changes from the Cepheid in the 
visible to the companion
in the ultraviolet. The locations of the spectrum centroids 
for the two stars
can be measured very accurately, better than 0.01" for the 
FOC. We have now
obtained first epoch observations of both AW Per and U Aql 
with the HST FOC
(Faint Object Camera) with the NUV (Near Ultraviolet) 
objective prism.

\section{Results} 

Measuring the inclination of a system requires two 
observations at phases
carefully selected from the ground-based orbit of the 
Cepheid. In addition, at
each epoch, several observations must be made with varying 
satellite roll angles
to determine the magnitude and orientation of the angular 
separation on the sky.

\begin{figure}
\begin{center}

\mbox{\epsfxsize=0.6\textwidth\epsfysize=0.6\textwidth\epsfbox{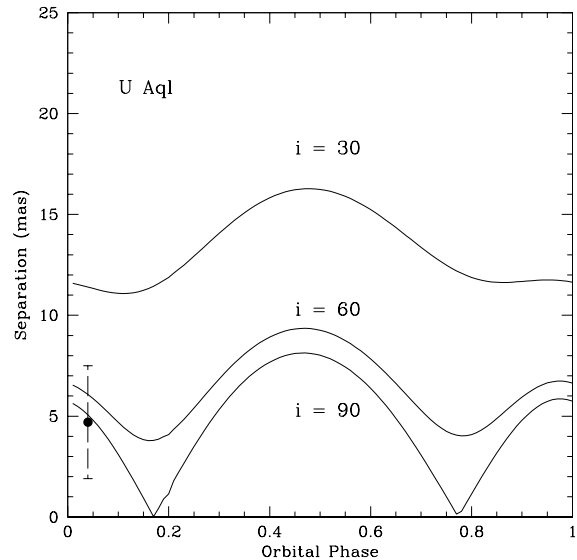}}
\caption{The predicted separation between the Cepheid U Aql 
A
and the companion (in mass) 
as a function of orbital phase from the ground-based 
spectroscopic orbit of the 
Cepheid. Three values of the inclination are illustrated. 
The point shows the value of the separation measured with 
the FOC. }
\end{center}
\end{figure}

Results are shown in Fig.\,1. The data are fitted to a 
model of the relative
contribution of each component based on observed IUE 
spectra of the composite
system, which, of course, includes the reddening (Evans, 
1992, 1994). (Residual
distortions in the field have been removed with a fourth 
order polynomial.) The
three values of the projected separation (``Amp'' in 
Fig.\,1), are then fitted
as a function of the roll angle to determine the full 
separation.

Using this approach, we have obtained a clear resolution of 
the Cepheid AW Per
at one epoch (separation 0.013 $\pm$ 0.003"). We also have 
a marginal
resolution of U Aql.

What can we say about the results so far\,? Fig.\,2 is an 
example using U\,Aql,
showing the predicted separations for three inclinations 
from the spectroscopic
orbit of the Cepheid (Welch et al., 1987). The mass ratio 
between the Cepheid
and the companion is taken from this orbit and the velocity 
of the companion
measured with the HST Goddard High Resolution Spectrograph 
(Evans et al., 1998).
While a second centroid measurement is required to 
determine the inclination
independently without using a mass ratio, Fig. 2 shows that 
combination of the
separation already measured and the mass ratio requires a 
high inclination.
Full results will be given in Massa \& Evans (2002).

\section{Future work} 

Observations for the second epoch are being scheduled 
on the HST STIS
spectrograph. STIS should provide improved accuracy because 
of easier
extraction from geometrically distinct pixels, and higher 
photometric
sensitivity.

\section*{Discussion}
\parindent=0pt

{\it M. Jerzykiewicz~:} How sensitive is your method to 
interstellar 
reddening? \medskip

{\it N. Evans~:} The model to which the centroid shift is 
compared is 
created from the observed IUE composite spectrum so it 
already
includes reddening. In addition, 
each star completely dominates at one end of the spectrum, 
so fluxes of each
star are well determined. \bigskip

{\it T. Bedding~:} Comment: The idea of using centroid 
variations as a 
function of wavelength from spectra has also been developed 
by Jeremy Bailey
from AAO who calls it ``spectro-astrometry'' (e.g. Bailey, 
1998, \mnras, 301,
161). Question: There seems to be a scatter in your 
luminosity-mass 
relation (e.g.\ Fig.\,5 in Evans et al., 1998). Could this 
be due to 
rotation as an extra parameter? \medskip

{\it N. Evans~:} Rotation is certainly a possibility. At 
this point, I'm 
looking forward to reducing the error bars enough to be 
able to make 
strong statements about something like that. \bigskip

\end{document}